\documentstyle[preprint,aps,epsfig]{revtex}  

\tighten       

\begin{document}
\draft 

\title{Asymptotic description of 
transients and synchronized states of globally coupled oscillators}
\author{ J.\ A.\ Acebr\'on \cite{acebron:email} and L.\ L.\ Bonilla 
\cite{bonilla:email}}
\address{Escuela Polit\'ecnica Superior,  Universidad Carlos III de Madrid,
Butarque 15, 28911 Legan{\'e}s, Spain}
\date{May 19, 1997}
\maketitle
\begin{abstract}
A two-time scale asymptotic method has been introduced to analyze 
the multimodal mean-field Kuramoto-Sakaguchi model of oscillator 
synchronization in the high-frequency limit. The method allows to
uncouple the probability density in different components corresponding
to the different peaks of the oscillator frequency distribution. Each
component evolves toward a stationary state in a comoving frame and the
overall order parameter can be reconstructed by combining them. Synchronized
phases are a combination of traveling waves and incoherent solutions 
depending on parameter values. Our results agree very well with direct
numerical simulations of the nonlinear Fokker-Planck equation for the 
probability density. Numerical results have been obtained by finite 
differences and a spectral method in the particular case of bimodal
(symmetric and asymmetric) frequency distribution with or without external
field. We also recover in a very easy and intuitive way the only other 
known analytical results: those corresponding to reflection-symmetric bimodal 
frequency distributions near bifurcation points.
\end{abstract}

\pacs{ }

\narrowtext

\section{Introduction}

In recent years mathematical modeling and analysis of synchronization
phenomena received increased attention because of its occurrence in quite
different fields, such as solid state physics \cite{TS1,TS2,CDW},
biological systems \cite{WIN,STRO,MISTRO,GRAY}, chemical reactions
\cite{KURAM2}, etc. These phenomena can be modeled in terms of populations
of interacting, nonlinearly coupled oscillators as first proposed by
Winfree \cite{WIN}. While the dynamic behavior of a small number of
oscillators can be quite interesting \cite{ARO}, here we are concerned 
with synchronization as a collective phenomenon for large populations 
of interacting oscillators \cite{STRO}. Then we can describe populations 
of oscillators interacting via simple couplings (e.g., all-to-all,
mean-field couplings) by means of kinetic equations for one-oscillator
densities~\cite{BON1,STRO,STROMI}. Many recent studies of synchronization 
phenomena combine numerical simulations with linear stability and bifurcation 
analyses of particular (stable) incoherent and synchronized 
states~\cite{STROMI,BNS,CRAW,OKUDA}. These works have described the onset 
of synchronized phases and, near degenerate bifurcation points, synchronized 
phases from their beginning to their end in the corresponding bifurcation 
diagram~\cite{BPVS}. In this paper we introduce a high-frequency singular 
perturbation method which describes (in a conveniently analytical manner) 
synchronized phases far from bifurcation points. The method nicely agrees 
with the results of numerical simulations. 

These ideas may be made concrete in a simple model put forth by Kuramoto 
and Sakaguchi \cite{KURAM,SAK} (see also \cite{STRO}). It consists of a 
population of coupled phase oscillators, $\theta_{j}(t)$, having natural 
frequencies $\omega_{j}$ distributed with a given probability density 
$g(\omega)$, and subject to the action of an external field $h_j$ 
which is distributed with a probability density $f(h)$,
\begin{equation}
  \dot{\theta_{j}} = \omega_{j} + \xi_{j}(t) - h_j\,\sin\theta_j 
   + \sum_{l=1}^{N} K_{jl} \sin(\theta_{l} -\theta_{j}),	\quad\quad
j=1,\ldots,N.
\label{1} 
\end{equation}
Here $\xi_{j}$ are independent white noise processes with expected values
\begin{equation}
	 \langle \xi_{j}(t) \rangle = 0,    \ \ \
      \langle \xi_{j}(t)\xi_{l}(t') \rangle = 2 D \delta(t -t')\,
			   \delta_{jl}.\label{2}
\end{equation}
In the absence of external field and white noise, each oscillator tries 
to run independently at its own frequency while the coupling tends to 
synchronize it to all the others. When the coupling is sufficiently weak, 
the oscillators run incoherently whereas beyond a certain threshold 
collective synchronization appears spontaneously. So far, several 
particular prescriptions for the matrix $K_{jl}$ have been considered. 
For instance, $K_{jl}=K>0$ only when $|j-l|=1$, and $K_{jl}=0$ otherwise 
(next-neighbor coupling) \cite{STROMI2}; $K_{jl}=K/N>0$ (mean-field 
coupling) \cite{KURAM,KURAM2}; hierarchical coupling \cite{LUMER}; 
random long-range coupling \cite{BON2,PAB,BON3} or even state 
dependent interactions \cite{SOMP}. In the mean-field case, the
model (\ref{1})-(\ref{2}) can be written in a convenient form by defining
the (complex-valued) order-parameter 
\begin{equation}
	    r e^{i\psi}=\frac{1}{N}\sum_{l=1}^{N} e^{i\theta_{l}}\,.
   \label{3}
\end{equation}
Here $|r(t)|$ measures the phase coherence of the oscillators, and
$\psi(t)$ measures the average phase. Then eq. (\ref{1}) reads
\begin{equation}
    \dot{\theta_{j}}=\omega_{j} - h_j \sin\theta_j + Kr\sin(\psi - 
     \theta_{j}) + \xi_{j}(t),  \ \ \	j=1, 2, \ldots, N.
  \label{4}
\end{equation}
In the limit of infinitely many oscillators, $N \rightarrow \infty$, a
{\it nonlinear Fokker-Planck equation} (NLFPE) was derived \cite{BON1,STROMI} 
for the one-oscillator probability density, $\rho(\theta,t,\omega,h)$,  
\begin{equation}
       \frac{\partial \rho}{\partial t}=D \frac{\partial^{2} \rho}{\partial 
\theta^{2}} -		\frac{\partial}{\partial \theta}(v \rho),\label{5}
\end{equation}
the drift-term being given by
\begin{equation}
	      v(\theta,t,\omega) = \omega - h \sin\theta + K r sin(\psi - \theta),
\label{6}
\end{equation}
and the order-parameter amplitude by
\begin{equation}
    r e^{i \psi} = \int_{0}^{2 \pi}\!\!\int_{-\infty}^{+\infty}
\!\!\int_{-\infty}^{+\infty}	e^{i\theta} \rho(\theta,t,\omega,h)\, g(\omega)\,
f(h)\, d\theta\, d\omega\, dh.
\label{7} 
\end{equation}
The probability density is required to be $2 \pi$-periodic as a function
of $\theta$ and normalized according to
\begin{equation}
       \int_{0}^{2 \pi} \rho(\theta,t,\omega,h)\, d\theta = 1. \label{8}
\end{equation}

Mean-field models such as those described above were studied, e.g., by
Strogatz and Mirollo \cite{STROMI} in the absence of external field and 
for a {\it unimodal} [$g(\omega)$ is non-increasing for $\omega>0$] 
frequency distribution, $g(\omega)$, having reflection symmetry, 
$g(-\omega) = g(\omega)$. In \cite{STROMI}, the authors showed that 
for $K$ smaller than a certain value $K_c$, the incoherent equiprobability 
distribution, $\rho_0 \equiv 1/(2\pi)$, is {\it linearly stable}, and 
linearly unstable for $K>K_c$. As $D\to 0+$, the incoherence solution 
is still unstable for $K>K_c$ [$=2/\pi g(0)$ at $D=0$], but it is 
neutrally stable for $K<K_c$: the whole spectrum of the equation 
linearized about $\rho_0$ collapses to the imaginary axis. In \cite{BNS}, 
the {\it nonlinear} stability issue was addressed, and the case of a 
reflection-symmetric {\it bimodal} frequency distribution was considered 
[$g(\omega)$ is even and it has maxima at $\omega = \pm \omega_0$]. 
In this case, new bifurcations appear, and bifurcating synchronized 
states have been asymptotically constructed in the neighborhood of 
the bifurcation values of the coupling strength. The {\it nonlinear} 
stability properties of such solutions were also studied for the 
explicit discrete example $g(\omega) = {1\over 2}[\delta(\omega-\omega_0) 
+ \delta(\omega+\omega_0)]$, cf.~\cite{BNS}. A complete bifurcation study 
taking into account the reflection symmetry of $g(\omega)$ was carried 
out by Crawford, \cite{CRAW}. Similar results were obtained by Okuda 
and Kuramoto in the related case of mutual entrainment between populations 
of coupled oscillators with different frequencies \cite{OKUDA}. 
Furthermore, a two-parameter bifurcation analysis near the tricritical 
point (at which bifurcating stationary and oscillatory solution 
branches coalesce) allows us to visualize a global bifurcation 
diagram in which oscillatory solution branches may be calculated 
analytically from their onset to their end~\cite{BPVS}. The effect of an 
external field on Kuramoto models has been analyzed in 
Refs.~\cite{ARE1,ARE2}. 

In this paper we shall illustrate our high-frequency perturbation 
method by applying it to the generalized mean-field Kuramoto model 
(\ref{5})-(\ref{8}). We shall assume that the frequency distribution is 
{\em multimodal} in the high-frequency limit: $g(\omega)$ has $m$ maxima
located at $\omega_0\Omega_l$, $l=1,\ldots,m$, where $\omega_0\rightarrow 
\infty$. Then $g(\omega)\, d\omega$ tends to the limit distribution 
\begin{eqnarray}
\Gamma(\Omega)\, d\Omega \equiv \sum_{l=1}^{m}\alpha_l\,\delta(\Omega 
-\Omega_l)\, d\Omega, \label{9}\\
\mbox{with}\quad\quad \sum_{l=1}^{m} \alpha_l = 1,\quad\mbox{and}\quad 
\Omega = \frac{\omega}{\omega_{0}}\,, \nonumber
\end{eqnarray}
independently of the shape of $g(\omega)$ as $\omega_0\to\infty$. 
Then $g(\omega)\, d\omega$ and $\Gamma(\Omega)\, d\Omega$ may be used
interchangeably when calculating any moment of the probability density 
[including of course the all-important order parameter (\ref{7})]. Thus any 
frequency distribution is equivalent to a discrete multimodal distribution 
in the high-frequency limit. The discrete symmetric bimodal distribution 
considered in \cite{BNS,BPVS} corresponds to $m=2$, $\Omega_l = (-1)^l$, 
$\alpha_l = {1\over 2}$, $l=1,2$. We shall show that the oscillator 
probability density splits into $m$ components, each contributing a wave 
rotating with frequency $\Omega_l \omega_0$ to the order parameter. 
The envelope of each component evolves to a stationary state as the 
time elapses. Thus our method yields analytical expressions for the 
probability density and the order parameter during the transients toward
synchronized (or incoherent) phases, which agree with direct numerical 
simulations of the NLFPE. Since it is not a small-amplitude expansion,
our method is valid well inside the regions of stable synchronized phases
in the phase diagram, far from bifurcation points. Of course we have derived
the method in the limit $\omega_0\to\infty$, but comparison with numerical 
simulations shows that $\omega_0=7$ is already close to infinity for 
all practical purposes. 

Our numerical calculations have been carried 
out by means of finite difference schemes and by using a spectral method
which generates a hierarchy of ordinary differential equations for
moments of the probability density which include the order parameter.
This method is equivalent to an expansion of the probability density in 
a Fourier series and it could in principle be used to reconstruct it.
The moment hierarchy was derived directly from Eqs.~(\ref{1})-(\ref{2})
by P\'erez Vicente and Ritort~\cite{PR}. They assumed that arithmetic means
converged to ensemble averages in the limit $N\to\infty$ [keeping $t=O(1)$], 
which was justified in~\cite{BON1}. From the moment hierarchy, P\'erez Vicente 
and Ritort~\cite{PR} also derived a nonlinear kinetic equation for a 
moment-generating function $\Upsilon(\theta,y,t) = \intÊe^{\omega y}\,\rho(
\theta,t;\omega)\, g(\omega) d\omega$, which is related to functional-equation 
formulations of Equilibrium Statistical Mechanics~\cite{KELLER} and fluid 
turbulence~\cite{HOPF,KRAI}. 

The rest of the paper consists of a description of our method of multiple 
scales in Section II, comparison with numerical results in Section III, and
our conclusions in Section IV.

\section{Method of multiple scales}
The high frequency limit of the NLFPE can be analyzed by means of a method 
of multiple scales. Let us change variables to a comoving frame and 
therefore rewrite the equation as 
\begin{eqnarray}
\frac{\partial \rho }{\partial t} &=& \frac{\partial^{2}\rho }{\partial 
\beta^{2}} - \frac{\partial U\, \rho }{\partial \beta }\,,
\label{mm1}\\
U &=& K r \sin(\psi - \beta - \omega t) - hÊ\sin(\beta + \omega t)\nonumber\\
&=& K r \sin\left(\psi - \beta - {\Omega\over\varepsilon} t\right) 
- hÊ\sin\left(\beta + {\Omega\over\varepsilon} t\right) \,,
\label{mm2}\\
\beta &=& \theta - \omega t\equiv \theta - {\Omega\over\varepsilon} t,  
\label{mm3}\\
\varepsilon &=& \frac{1}{\omega_{0}}\ll 1.
\label{mm4}
\end{eqnarray}
The order parameter is now
\begin{equation}
re^{i\psi }=\sum_{j=1}^m \alpha_j e^{i\Omega_{j}{t\over\varepsilon}}\int 
e^{i\beta }\,\rho(\beta,t;\Omega_j,h;\varepsilon)\, f(h)\, d\beta \, dh,  
\label{mm5}
\end{equation}
where we have used (\ref{9}) and have implicitly assumed that h=O(1). 
The discrete character of the frequency distribution in the high-frequency
limit makes it possible to simplify (\ref{mm1}). In fact, Eq.~(\ref{mm5}) 
shows that $\rho$ may be split in different components $\rho_j = 
\rho(\beta,t;\Omega_j,h;\varepsilon)$. Therefore we can write (\ref{mm1}) 
as a coupled system of equations for the density components $\rho_j$:
\begin{eqnarray}
\frac{\partial \rho_{j}}{\partial t} &=& \frac{\partial^{2}\rho_{j}}{\partial 
\beta^{2}} - \frac{\partial (U_{j}\, \rho_{j})}{\partial \beta}\,,
\label{mm6}\\
U_{j} &=& \mbox{Im}\left\{ K\,\sum_{l=1}^m \alpha_l e^{i(\Omega_{l}-
\Omega_{j})\, {t\over\varepsilon}} 
\int e^{i(\beta'-\beta)}\, \rho(\beta,t;\Omega_l,h;\varepsilon)\, f(h)\, 
d\beta' dh  - h\, e^{i(\beta + \Omega_{j} {t\over\varepsilon})}
\right\} \,,
\label{mm7}\\
&& \int_0^{2\pi} \rho_j(\beta,t;h;\varepsilon)\, d\beta = 1. \label{norm}
\end{eqnarray}

Eqs.~(\ref{mm6}) and (\ref{mm7}) contain terms with rapidly-varying 
coefficients. It is then to be expected that an appropriate 
asymptotic method will be able to average them out thereby capturing
the slow evolution of $\rho_j$ (or perhaps its envelope). This may be
achieved by introducing fast and slow time scales as follows: 
\begin{eqnarray}
\tau = \frac{t}{\varepsilon},\quad\quad t=t.  \label{mm8}
\end{eqnarray}
We look for a distribution function which is a 2$\pi$-periodic function 
of $\beta$ according to the Ansatz: 
\begin{eqnarray}
\rho(\beta,t;\Omega,h;\varepsilon) = \rho^{(0)}(\beta,\tau,t;\Omega,h)
+ \varepsilon \rho ^{(1)}(\beta,\tau,t;\Omega,h) + 
O(\varepsilon^2),  \label{mm9}
\end{eqnarray}
Inserting (\ref{mm9}) into (\ref{mm6})-(\ref{mm7}), we obtain the 
following hierarchy of equations

\begin{eqnarray}
{\partial\rho_{j}^{(0)}\over\partial\tau } &=& 0,  \label{mm11}\\
{\partial\rho_{j}^{(1)}\over\partial\tau } &=& -\frac{\partial}{\partial \beta} 
\left\{ \rho_{j}^{(0)}\,\mbox{Im} \,\left( K \sum_{l\neq j} \alpha_j \, 
e^{i(\Omega_{l} - \Omega_j) \tau}\, e^{-i\beta} Z_l^{(0)}) - h \, e^{i(\beta 
+\Omega_j \tau)}) \right) \right\}\nonumber\\
&-& {\partial\rho_{j}^{(0)}\over \partial t} + D {\partial^{2}\rho_{j}^{(0)}
\over\partial\beta^{2}} 
-  K \alpha_{j} \,
\frac{\partial}{\partial \beta} \left\{\rho_{j}^{(0)}\, \mbox{Im}\,
(e^{-i \beta} Z_{j}^{(0)})\right\} \,,
\label{mm12}
\end{eqnarray}
where
\begin{equation}
Z_j^{(0)}(t) = \int e^{i\eta} \rho_j^{(0)}(\eta,t,h)\, f(h)\, d\eta\, dh.    
\label{mm13}
\end{equation}

Eq.~(\ref{mm11}) implies that  $\rho_{j}^{(0)}$ is independent of $\tau$. 
Then the terms in the right side of (\ref{mm12}) which do not have
$\tau$-dependent coefficients give rise to secular terms (unbounded on 
the $\tau$-time scale). The condition that no secular terms should  
appear is
\begin{equation}
{\partial\rho_{j}^{(0)}\over \partial t} - D {\partial^{2}\rho_{j}^{(0)}
\over\partial\beta^{2}} + K \alpha_{j} \,\frac{\partial}{\partial \beta} 
\left\{ \rho_{j}^{(0)}\, \mbox{Im}\, (e^{-i \beta} Z_{j}^{(0)}) \right\} 
= 0.   \label{mm14}
\end{equation}
This equation should be solved for $\rho_j^{(0)}$ together with (\ref{mm13}), 
the normalization condition 
\begin{eqnarray}
\int_0^{2\pi} \rho_j^{(0)}(\beta,t;h)\, d\beta = 1, \label{norm_0}
\end{eqnarray}
and an appropriate initial condition. We see that, except for the 
$h$-integration in (\ref{mm13}), this problem is equivalent 
to solving a NLFPE with frequency distribution $g(\omega)=\delta(\omega)$, 
(identical oscillators) and coupling constant $K_j = K\alpha_j$. If the 
initial condition is independent of the external field $h$, we know that 
the solution of the previous NLFPE evolves towards a stationary state as
time elapses~\cite{STROMI}. If the initial condition depends on $h$, all 
we can say is that $\int \rho_j^{(0)}(\beta,t;h)\, f(h)\, dh$ tends to 
a stationary state independent of $h$ as $t\to\infty$. In both cases all 
possible stationary states are solutions of Eqs.~(\ref{mm15})-(\ref{mm16}) 
below~\cite{BNS}
\begin{eqnarray}
\rho_j^{(0)}(\beta) = \frac{e^{K\alpha_{j}R_{j}D^{-1}\cos(\Psi_{j}-\beta)}
\,\int_{0}^{2\pi}d\beta'\, e^{-K\alpha_{j}R_{j}D^{-1}\cos(\Psi_{j}
-\beta-\beta')}}{\int_{0}^{2\pi} d\beta\, e^{K\alpha_{j}R_{j}D^{-1}
\cos(\Psi_{j}-\beta)}\,\int_{0}^{2\pi}d\beta'\, e^{-K\alpha_{j}R_{j}D^{-1}
\cos(\Psi_{j}-\beta-\beta')}}\,.\label{mm15}
\end{eqnarray}
The order parameter $R_j e^{i\Psi_{j}}$ is calculated by inserting (\ref{mm15}) 
into (\ref{mm13}):
\begin{equation}
 R_j e^{i\Psi_{j}} = \int_0^{2\pi} e^{i\eta} \rho_j^{(0)}(\eta)\, d\eta 
\equiv  \lim_{t\to\infty} Z_j^{(0)}(t). 
\label{mm16}
\end{equation}
For $K\alpha_j < 2D$, the only stationary solution is $\rho_j^{(0)} = 
{1\over 2\pi}$ (incoherence), which is stable. At $K\alpha_j = 2D$, 
a stable branch of synchronized solutions bifurcates supercritically 
from incoherence. They exist for all $K\alpha_j > 2D$. 

The overall order parameter (\ref{mm5}) is given by
\begin{equation}
r e^{i\psi} = \sum_{j=1}^m \alpha_j \, R_j \, e^{i(\Omega_{j}\tau +
\Psi_{j})} + O(\varepsilon)\,. 
\label{mm17}
\end{equation}
To find $\psi$, we multiply both sides of (\ref{mm17}) by $e^{-i\psi}$,
and then take imaginary and real parts. After a little algebra, we obtain
\begin{eqnarray}
\tan \psi = \frac{\sum_{j=1}^{m}\alpha_{j}\, R_{j}\,\sin(\Omega_{j}\tau
+\Psi_{j})}{\sum_{j=1}^{m}\alpha_{j}\, R_{j}\,\cos(\Omega_{j}\tau
+\Psi_{j})}\,,\label{mm18}\\
r = \sum_{j=1}^{m}\alpha_{j}\, R_{j}\,\cos(\Omega_{j}\tau
+\Psi_{j} - \psi)\,.
\label{mm19}
\end{eqnarray}
Notice that $r$ in (\ref{mm19}) may be negative, positive or zero. 
Then the amplitude of the overall order parameter is $|r(t)|$.

Let us now consider, for the sake of definiteness, the special case of 
an asymmetric bimodal frequency distribution, with zero external field, 
\begin{eqnarray}
\Gamma(\omega) = \alpha\, \delta(\Omega-1) + (1-\alpha)\, \delta(\Omega +
1),  \quad 0< \alpha <1, \quad\quad f(h) = \delta(h),
\end{eqnarray}
and analyze the possible synchronized states. Eqs.~(\ref{mm18}) and 
(\ref{mm19}) become 
\begin{eqnarray}
\tan \psi =\frac{\alpha\, R_{+}\,\sin(\Psi_{+} + \tau) + (1-\alpha)\, R_{-}\,
\sin(\Psi_{-} - \tau)}{\alpha\, R_{+}\,\cos(\Psi_{+} + \tau) + (1-\alpha)\, 
R_{-}\, \cos(\Psi_{-} - \tau)}\,,
\label{mm20}\\
r= \alpha R_+ \cos(\Psi_+ + \tau -\psi) + (1-\alpha) R_- \cos(\Psi_- - \tau
-\psi).
\label{mm21}
\end{eqnarray}
Let us now assume that $\alpha < 1/2$ to be specific. Then we have the 
following possibilities depending on the value of the coupling constant:
\begin{enumerate} 
\item If $0<K< 2D/(1-\alpha)$, the incoherent solution $\rho = 1/(2\pi)$
is stable and it is the only possible stationary solution. 
\item If $2D/(1-\alpha) < K < 2D/\alpha$, a globally stable partially  
synchronized solution issues forth from incoherence at $K=2D/(1-\alpha)$. 
It has $R_+ = 0$, $\psi = \Psi_- - \tau$, and $r = (1-\alpha)\, R_-$. 
Its component $\rho_+ = 1/(2\pi)$ is incoherent while its component 
$\rho_-$ is synchronized according to Eq.~(\ref{mm15}). The overall 
effect is having a traveling wave solution (rotating clockwisely), 
once the angular variable $\beta$ is changed back to $\theta$ according 
to (\ref{mm3}). 
\item If $K> 2D/\alpha$, the component $\rho_+$ becomes partially 
synchronized too. The probability density then has traveling wave
components rotating clockwisely and anticlockwisely. Their order 
parameters have different strengths and $R_- >R_+$ if $\alpha<1/2$. 
\end{enumerate}

When $\alpha = 1/2$, both traveling wave components appear at the same 
value of the coupling constant, $K=4D$, and have equal strength:
$R_+ = R_- \equiv R$, $\Psi_+ = \Psi_- \equiv \Psi$. $R$ is the amplitude
of the order parameter corresponding to a unimodal frequency distribution
and a coupling constant $K_+ = K_- = \alpha K = K/2$. Then (\ref{mm20}) 
and (\ref{mm21}) imply that $\psi = \Psi + q\pi$ ($q$ is an integer number) 
and $r = R\,\cos(\omega_0 t + q\pi)$, respectively. Thus we have obtained  
an overall standing wave which is stable. Of course other possible 
solutions are traveling waves with $ R_{+}>0$, $R_{-} = 0$ and $R_{+} 
= 0$, $R_{-}>0$, which should be unstable because incoherence is an 
unstable solution of (\ref{mm15}) for the corresponding stationary
component $\rho_j$ if $K/2 > 2D$. These results coincide perfectly with 
those obtained by means of bifurcation theory in \cite{BPVS} and 
\cite{CRAW} for a symmetric bimodal frequency distribution ($\alpha 
= 1/2$). To see this, we recall that the stable (up to a constant shift 
in the origin of time which depends on initial conditions) standing wave 
probability density may be approximated near a bifurcation point $K_c= 4D$ 
by the following expressions~\cite{BPVS}:
\begin{eqnarray}
\rho(\theta,t,\omega) = {1\over 2\pi}\, [1 + \epsilon\,\sigma_1
+ O(\epsilon^2)],\quad\quad K = 4D + \epsilon^2 \,K_2,
\label{mm22}\\
\sigma_1 = A\, \left\{ \frac{e^{i(\Omega t + \theta)}}{D+ i (\Omega + 
\omega)} + \frac{e^{i(\Omega t - \theta)}}{D+ i (\Omega - \omega)} 
\right\} + cc,\quad\quad \Omega = \sqrt{\omega_{0}^{2} - D^{2}} ,
\label{mm23}	\\
A = \sqrt{\frac{\mbox{Re} \
	    \alpha}{\mbox{Re} \ (\gamma + \beta)}} \ e^{i\nu\epsilon^{2} K_{2}t},
     \ \ \   \nu = \mbox{Im} \ \alpha - \frac{\mbox{Im} \ (\gamma +
	 \beta)}{\mbox{Re} \ (\gamma + \beta)} \,\mbox{Re} \ \alpha \,,
\label{mm24}
\end{eqnarray}
where $cc$ means complex conjugate of the preceding term. As $\omega_0
\to + \infty$, the parameters $\alpha$, $\beta$, $\gamma$ of (\ref{mm24}) 
become~\cite{BPVS} 
\begin{eqnarray}
\alpha = {1\over 4}\,, \quad \beta = 0\,, \quad \gamma = {1\over 2 D 
K_{2}}\,.
\label{mm25}
\end{eqnarray}
Inserting (\ref{mm22}) to (\ref{mm25}) in Eq.~(\ref{7}) for the order 
parameter, we obtain that $\psi$ is constant and 
\begin{eqnarray}
r(t) = \sqrt{K - 4D\over 2D}\,\cos\omega_0 t + O((K - K_{c}))\,.
\label{mm26}
\end{eqnarray}

Now we can compare Eq.~(\ref{mm26}) with the result of our two-time scale
method, $r = R\,\cos\omega_0 t$. $R$ is the amplitude
of the order parameter corresponding to a unimodal frequency distribution
and a coupling constant $K_+ = K_- = \alpha K = K/2$. Near the bifurcation 
point, Eq.~(2.12) of Ref.~\cite{BNS} with $K_c = 2D$ (corresponding to 
$\omega_0 = 0$) and $K_{\pm} = K/2$ yields $R \sim \sqrt{(K/2 - 2 D)/D}$,
which implies exactly the result (\ref{mm26}). It is immediate to show 
that both methods also lead to the same expressions for traveling wave
solutions. 

\section{Numerical results}
\subsection{Spectral numerical method}
\begin{figure}
\centerline{\hbox{\psfig{figure=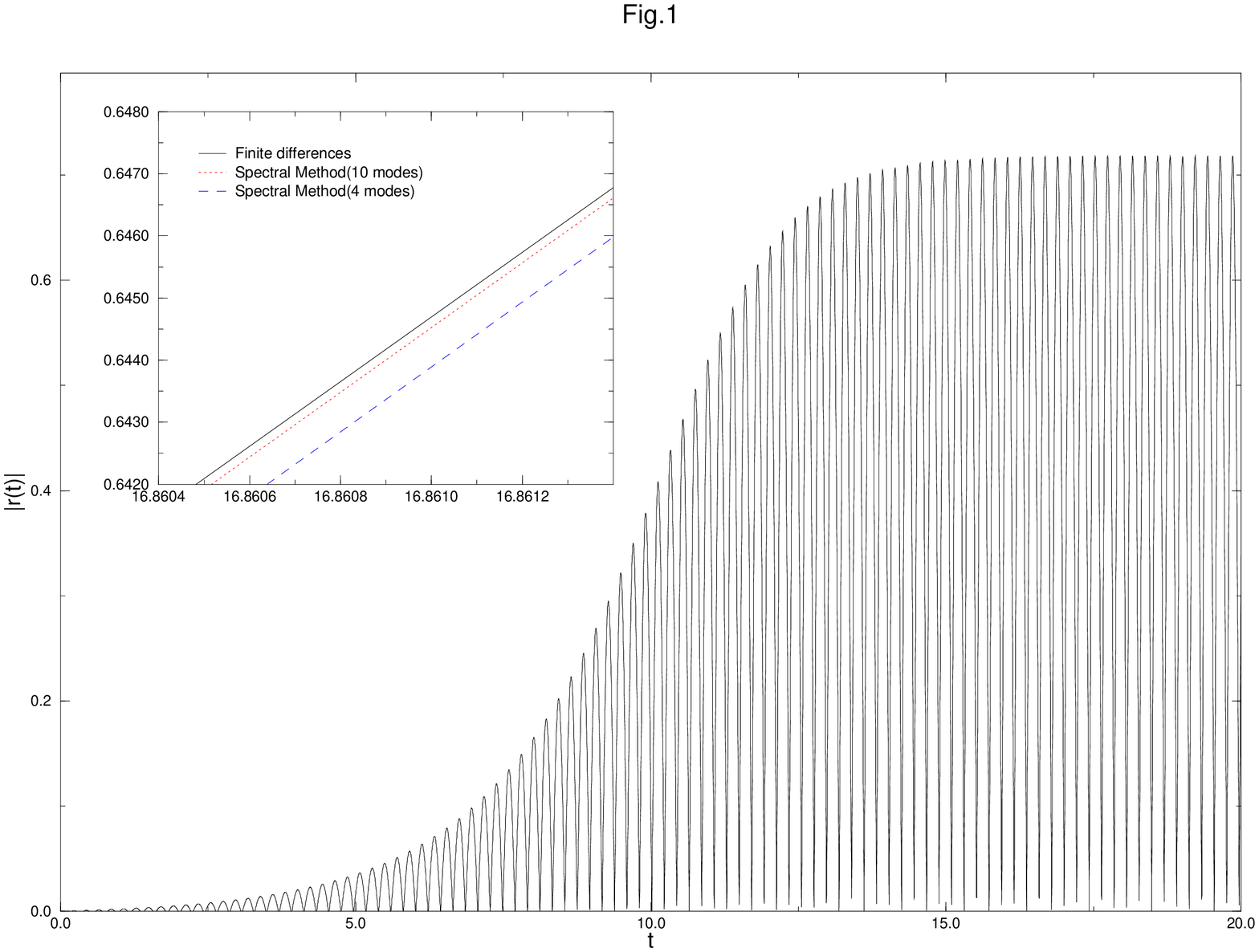,width=7.0cm,angle=-90}}}
\caption{Comparison between the results of numerical simulations by 
finite differences and our spectral method. We have a discrete bimodal 
frequency distribution of the oscillators, no external field and the 
following parameter values: $\omega_{0} = 15$, $K=6$ and $\alpha = 0.5$ 
(frequency distribution with reflection symmetry). Differences between the
methods are appreciated only on a rather fine time scale as the inset shows.
}
\label{fig1}
\end{figure}

Direct numerical simulations of the nonlinear Fokker-Planck system 
confirms our asymptotic results. We have studied discrete bimodal 
frequency distributions only, and used two different numerical
methods. A standard finite differences method may be used to numerically 
integrate (\ref{5}) - (\ref{8}) without stability problems up to 
frequencies $\omega_0 = 15$ (we set $D=1$ in all our computations). 
For larger frequency values, time steps below 0.008 were needed and the 
computing cost makes this method unpractical. As indicated in the previous
Section, the drift term dominates diffusion at higher frequency and 
the system acquires a quasi-hyperbolic character. To simulate the NLFPE
at high frequencies, we propose a simple spectral method, which we will 
describe in the simple case of $f(h) = \delta(h)$. The idea is 
to find a set of ordinary differential equations for {\em moments}
of the probability density related to the order parameter $r\, e^{i\psi}$: 
\begin{eqnarray}
x_{\pm}^{(j)} = \int_0^{2\pi} \rho(\theta,t,\pm\omega_0)\,\cos[j(\psi 
-\theta)]\, d\theta ,
\label{n1}\\
y_{\pm}^{(j)} = \int_0^{2\pi} \rho(\theta,t,\pm\omega_0)\,\sin[j(\psi 
-\theta)]\, d\theta ,
\label{n2}\\
r = \alpha\, x_+^{(1)} + (1-\alpha)\,  x_-^{(1)}\,.
\label{n3}
\end{eqnarray}
An infinite hierarchy of equations for these moments may be obtained
by differentiating (\ref{n1}) and (\ref{n2}) with respect to time
and then using the NLFPE and integration by parts to simplify the 
result. We obtain 
\begin{eqnarray}
{dx_{\pm}^{(j)}\over dt} = - j^2\, x_{\pm}^{(j)} \pm j \omega_0\,
y_{\pm}^{(j)} + {Kj\over 2}\, r x_{\pm}^{(j-1)} - {Kj\over 2}\, r 
x_{\pm}^{(j+1)} - j\, {d\psi\over dt}\, y_{\pm}^{(j)}\,,
\label{n4}\\
{dy_{\pm}^{(j)}\over dt} = - j^2\, y_{\pm}^{(j)} \mp j \omega_0\,
x_{\pm}^{(j)} + {Kj\over 2}\, r y_{\pm}^{(j-1)} - {Kj\over 2}\, r 
y_{\pm}^{(j+1)} + j\, {d\psi\over dt}\, x_{\pm}^{(j)}\,,
\label{n5}\\
r\, {d\psi\over dt} = \omega_0\, [\alpha\, x_{+}^{(1)} - (1-\alpha)\, 
x_{-}^{(1)}],
\label{n6}
\end{eqnarray}
As explained in the Introduction, an equivalent hierarchy may be derived 
directly from the Langevin equations (\ref{1}) and (\ref{2}) \cite{PR}. 
The numerical method consists of solving (\ref{n3})-(\ref{n6}) 
for $j = 1,\ldots, N$, with $x_{\pm}^{(N+1)} = y_{\pm}^{(N+1)} = 0$. 
The number of modes, $N$, should be chosen so large that the numerical 
results {\em for the order parameter} do not depend on it. A practical case is 
presented in Fig.~\ref{fig1} for $\omega_0 = 15$, $K=6$ and $\alpha = 0.5$ 
for which the method of finite differences is still practical. We see 
that keeping four modes ($N=4$) yields already quite good agreement. 
Let us now describe the results of our numerical simulations.

\subsection{Results for bimodal frequency distributions and no external field}
\begin{figure}
\centerline{\hbox{\psfig{figure=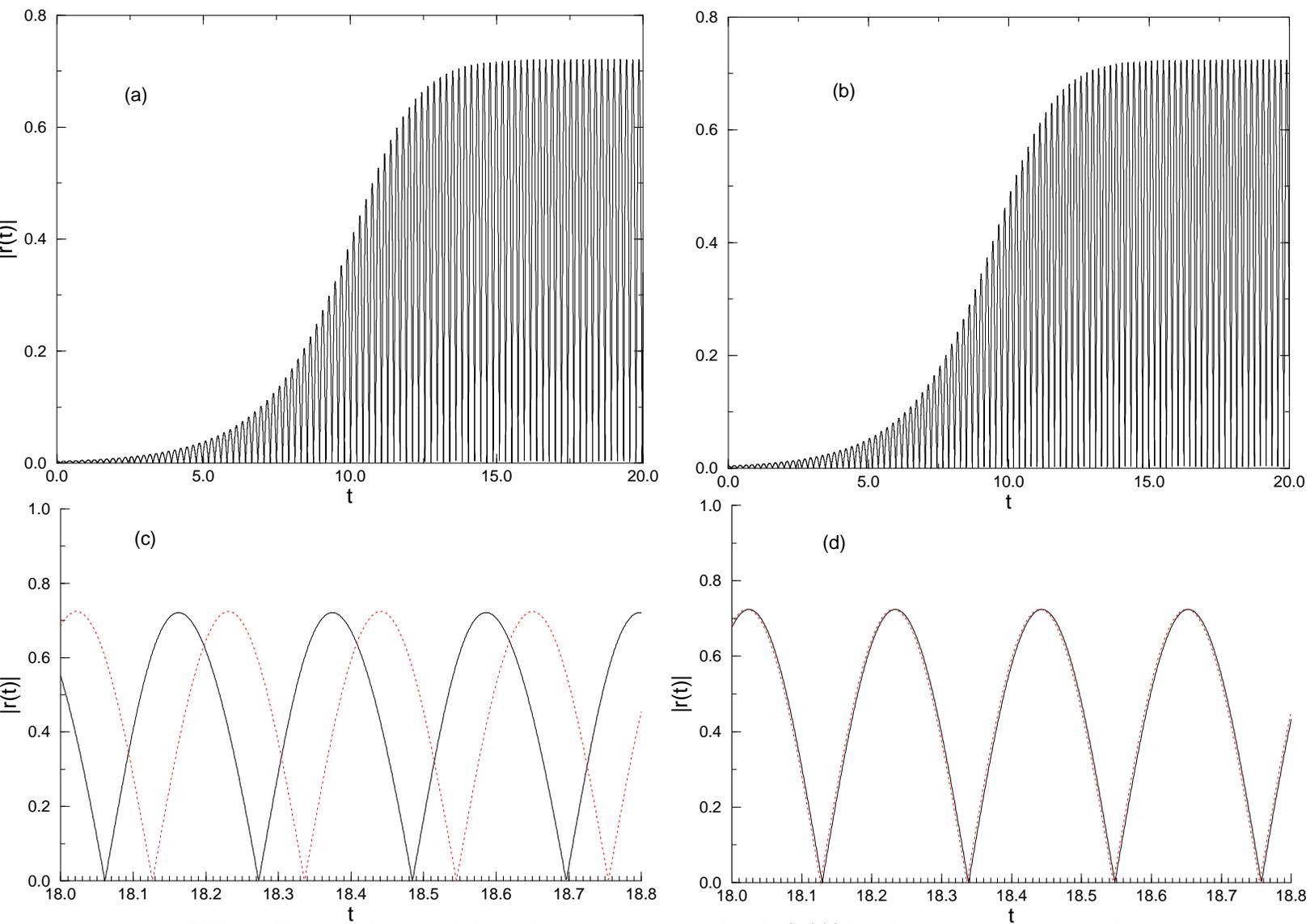,width=15.0cm,angle=-90}}}
\caption{Time evolution of the order parameter amplitude $|r(t)|$ for 
the same parameter values of Figure 1: (a) analytical result from our 
leading-order asymptotics; (b) numerical simulation; (c) comparison 
between both results; (d) same as (c) but now we have shifted the 
analytical result so that $t\to t+0.055$.}
\label{fig2}
\end{figure}

\begin{figure}
\centerline{\hbox{\psfig{figure=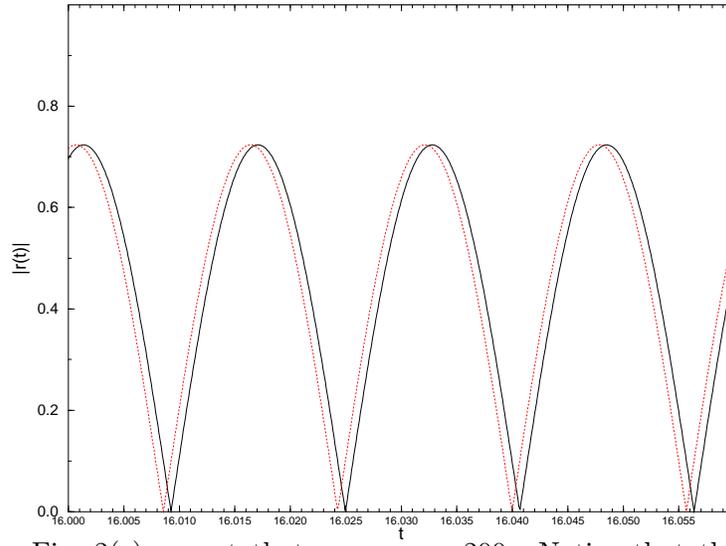,width=8.0cm,angle=-90}}}
\caption{Same as Fig.~2(c) except that now $\omega_0 = 200$. Notice that 
the time shift between analytical and numerical results is now much smaller. }
\label{fig3}
\end{figure}

\begin{figure}
\centerline{\hbox{\psfig{figure=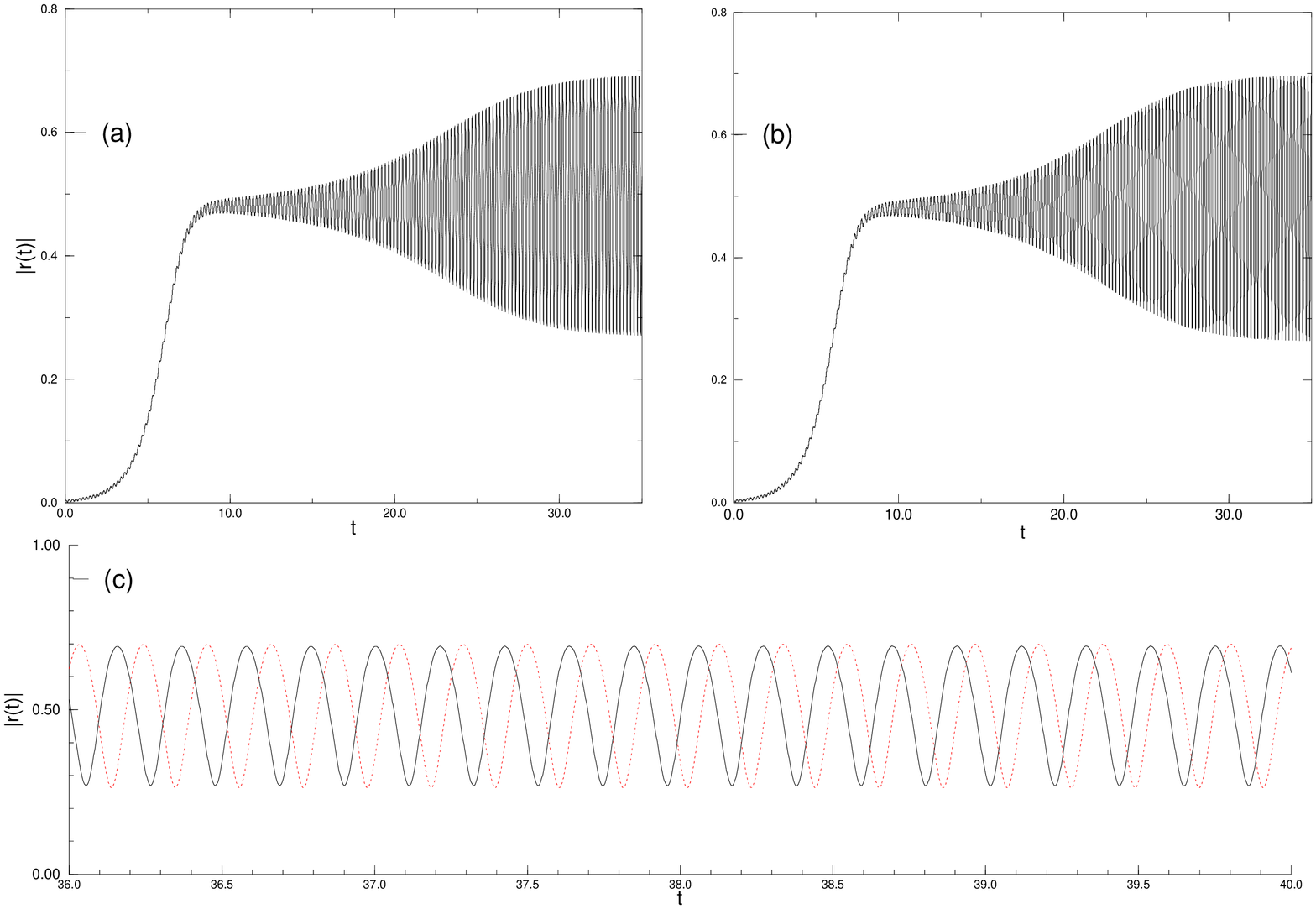,width=15.0cm,angle=-90}}}
\centerline{\hbox{\psfig{figure=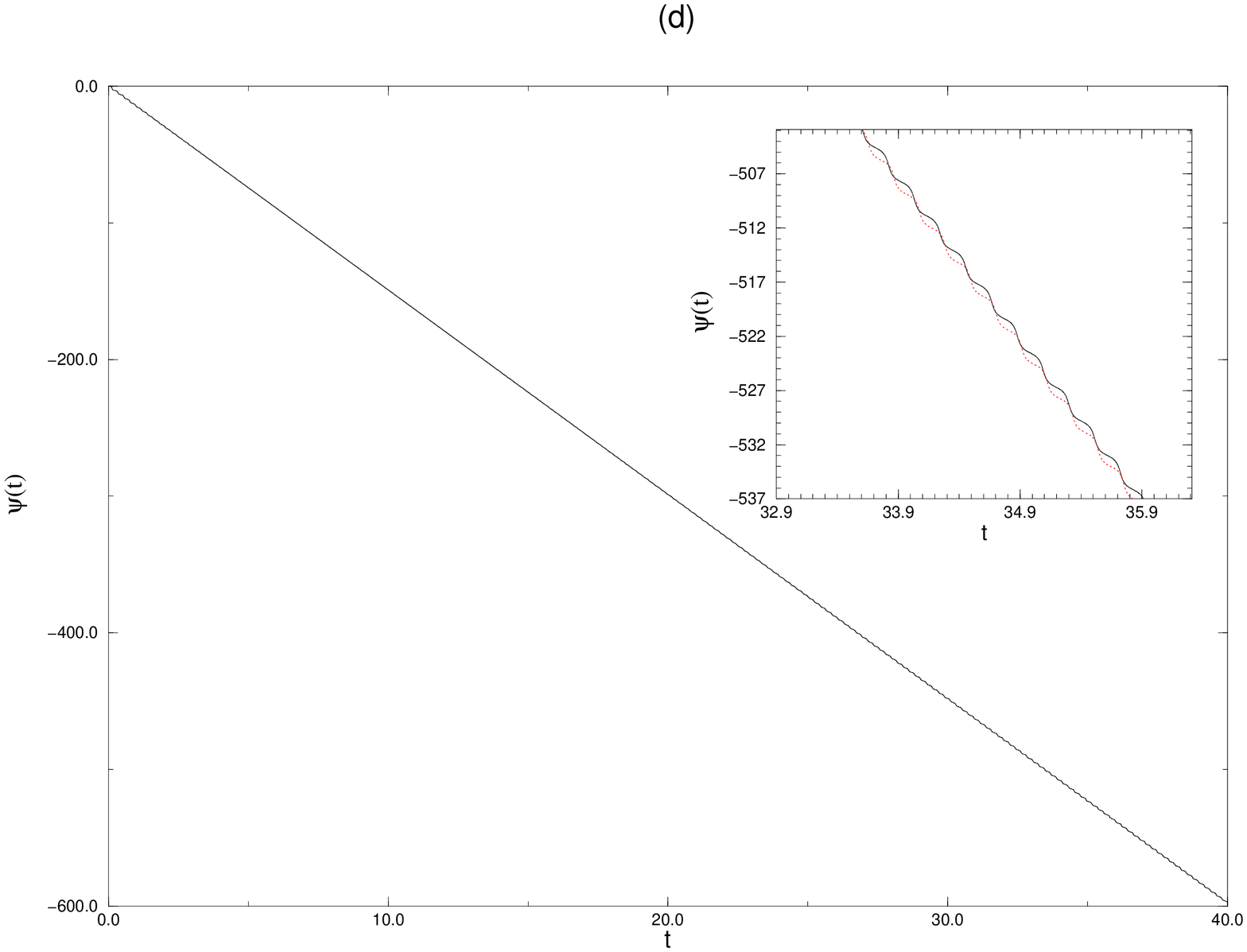,width=8.0cm,angle=-90}}}
\caption{Time evolution of the order parameter $r(t)\, e^{i\psi(t)}$ for an 
asymmetric bimodal frequency distribution. Parameters are the same as in 
Fig.~1, except that now $\alpha = 0.4$. (a) Analytical results for the 
evolution of $|r(t)|$; (b) numerical results; (c) comparison between both 
results; (d) evolution of the phase of the order parameter $\psi(t)$: there 
is only a small time shift between analytical and numerical results. 
Notice that for an asymmetric frequency distribution with these parameter 
values, $K>2/\alpha = 10/3$, so that the synchronized phase is an asymmetric 
combination of clockwise and anticlockwise rotating traveling waves.
}
\label{fig4}
\end{figure}

\begin{figure}
\centerline{\hbox{\psfig{figure=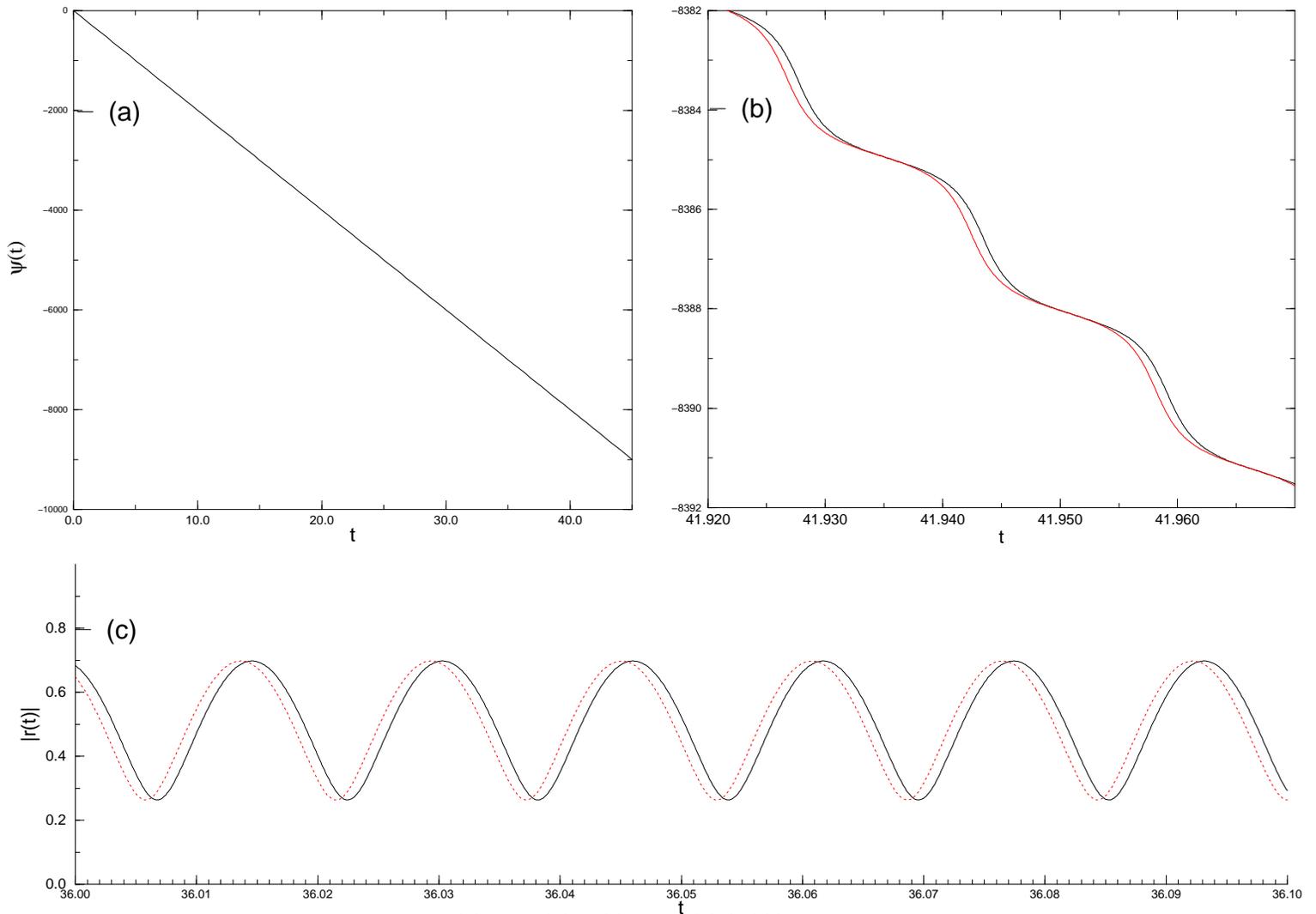,width=15.0cm,angle=-90}}}
\caption{Time evolution of (a) and (b) $\psi(t)$, and (c) $|r(t)|$,  for the 
asymmetric frequency distribution of Fig.~4 when $\omega_0 = 200$.
The other parameter values are as in Fig.~4.}
\label{fig5}
\end{figure}

\begin{figure}
\centerline{\hbox{\psfig{figure=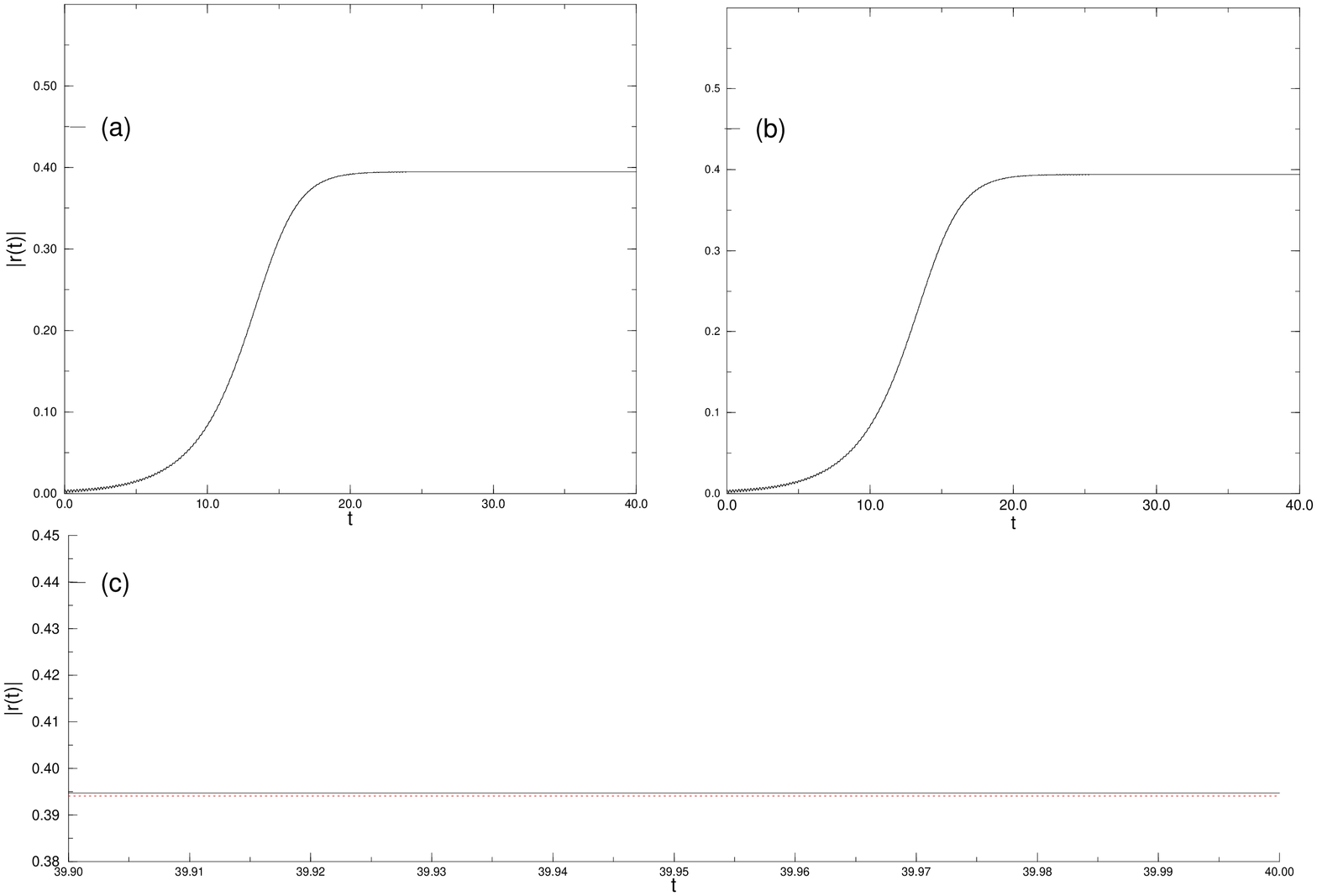,width=10.0cm,angle=-90}}}
\centerline{\hbox{\psfig{figure=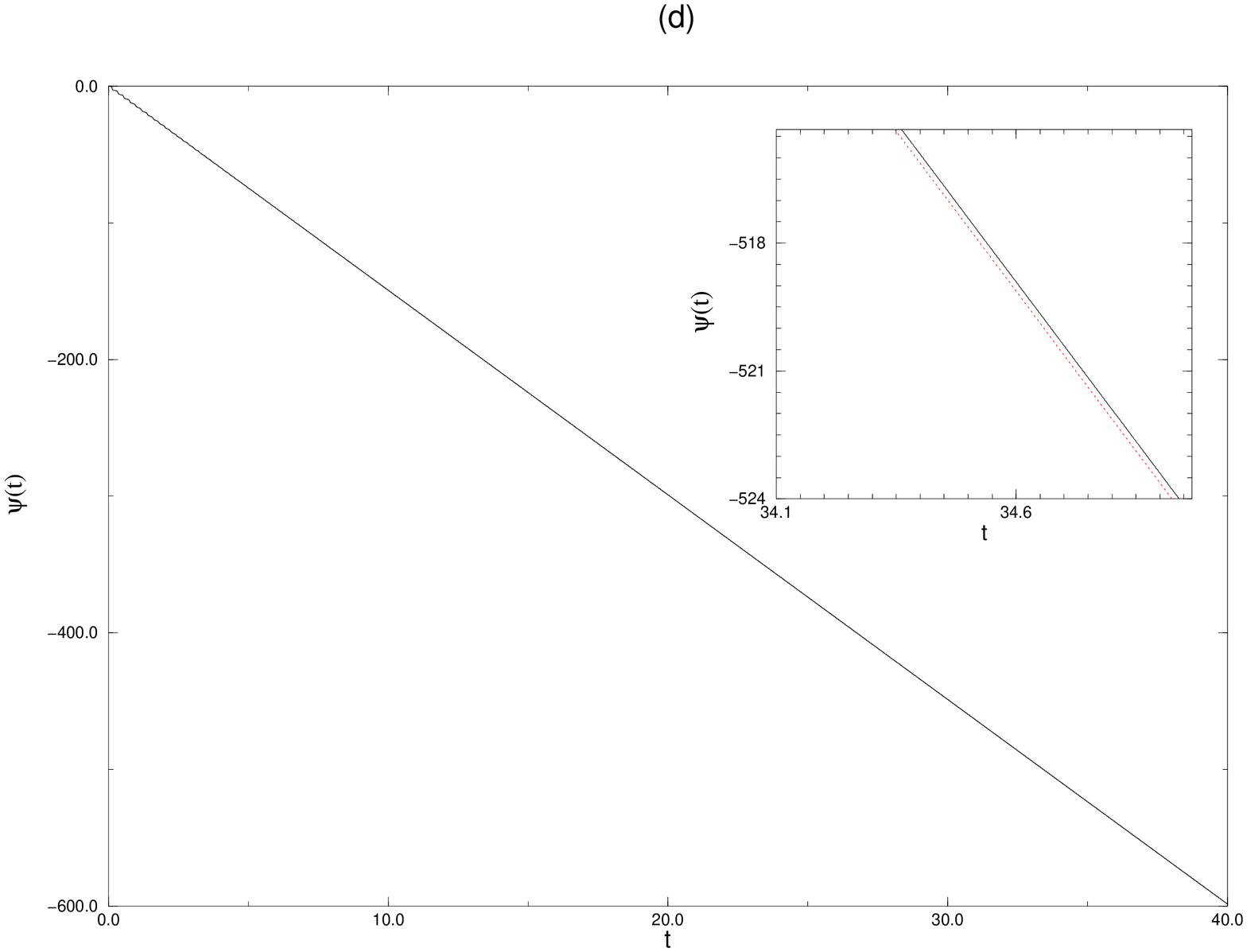,width=8.0cm,angle=-90}}}
\caption{Same as Fig.~4 for a lower value of the coupling constant, $K =
4.5$. Now $K\alpha = 1.8 < 2 < 2.7 = K(1-\alpha)$. Only the component of
the probability density with negative frequency is synchronized. Then we 
obtain a traveling wave rotating clockwisely with constant $|r(t)|$ and 
phase $\psi (t) = - \omega_0 t$.}
\label{fig6}
\end{figure}

We see in Figures \ref{fig2} to \ref{fig6} that our analytical (asymptotic) 
and numerical results agree very well except for a constant phase shift 
which decreases as $\omega_0$ increases (compare figures \ref{fig2} and 
\ref{fig3} corresponding to $\omega_0 = 15$ and 200, respectively). Results 
for an asymmetric bimodal frequency distribution without external field are 
depicted in Figs.~\ref{fig4} to \ref{fig6}. As explained in the previous 
Section, we obtain different synchronized phases depending on the value
of the coupling constant for each component of the probability density. 
In Figures \ref{fig4} and \ref{fig5}, $K>2D/\alpha>2D/(1-\alpha)$. Then 
each component of the probability density evolves towards a synchronized
phase rotating with its own frequency, $\pm\omega_0$, and with a 
constant amplitude of the order parameter given by the stationary 
expression (\ref{mm16}). The overall order parameter is given by 
Eqs.~(\ref{mm20})-(\ref{mm21}) and the difference between analytical 
and numerical results is a constant shift in time which diminishes as
the frequency $\omega_0$ increases. In Fig.~\ref{fig6} we observe the
situation for a smaller coupling constant such that only one density 
component is synchronized. We obtain a traveling wave whose order
parameter has a constant amplitude and a phase linearly decreasing 
with time. What happens if the frequency distribution has reflection
symmetry ($\alpha = 0.5$) is obvious: both density components have 
equal strength and therefore the phase of the order parameter is 
constant and its amplitude oscillates giving rise to a standing wave.
This is exactly what bifurcation theory predicts~\cite{CRAW,BPVS}. We
have checked the excellent agreement between our present asymptotic 
theory, the leading-order expression for the order parameter obtained by 
bifurcation theory, and direct numerical simulations. The results obtained
by these three methods are indistinguishable for $K=4.005$ ($K_c = 4$).

\subsection{Results for unimodal frequency distributions and deterministic 
external field} 
\begin{figure}
\centerline{\hbox{\psfig{figure=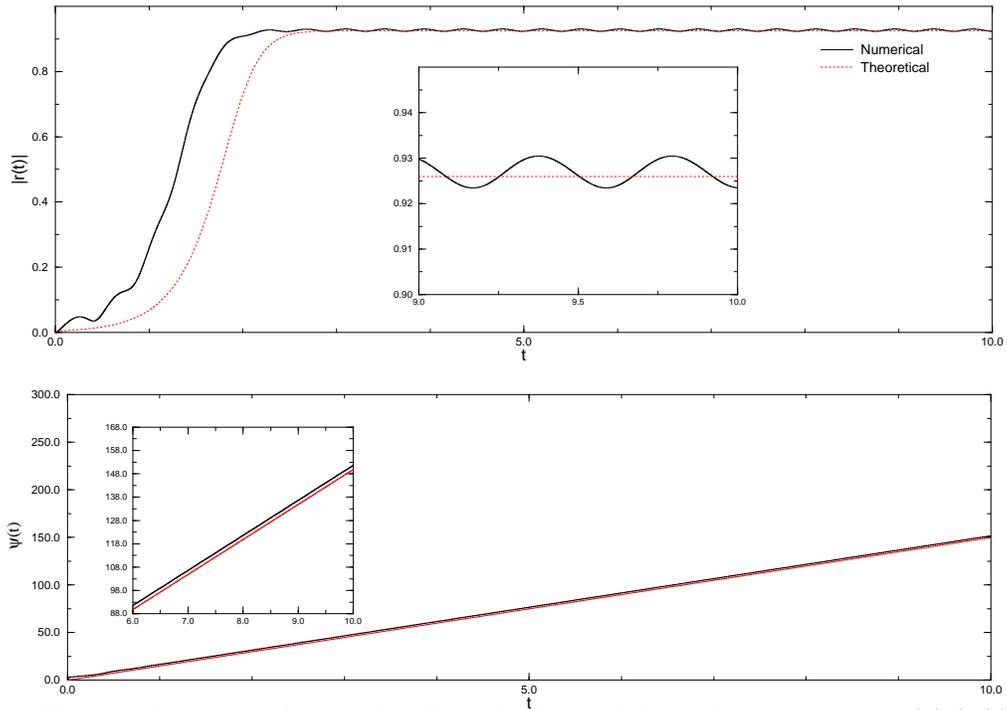,width=10.0cm,angle=-90}}}
\caption{Time evolution of the amplitude and phase of the order parameter:
(a) $|r(t)|$ and (b) $\psi(t)$, for the unimodal Kuramoto-Sakaguchi model. 
Parameter values are: $\omega_0 = 15$, $h_0 = 0.5$, and $K=7.5$. Notice 
the additional oscillation of the amplitude which is not predicted by
our leading-order asymptotics.}
\label{fig7}
\end{figure}

\begin{figure}
\centerline{\hbox{\psfig{figure=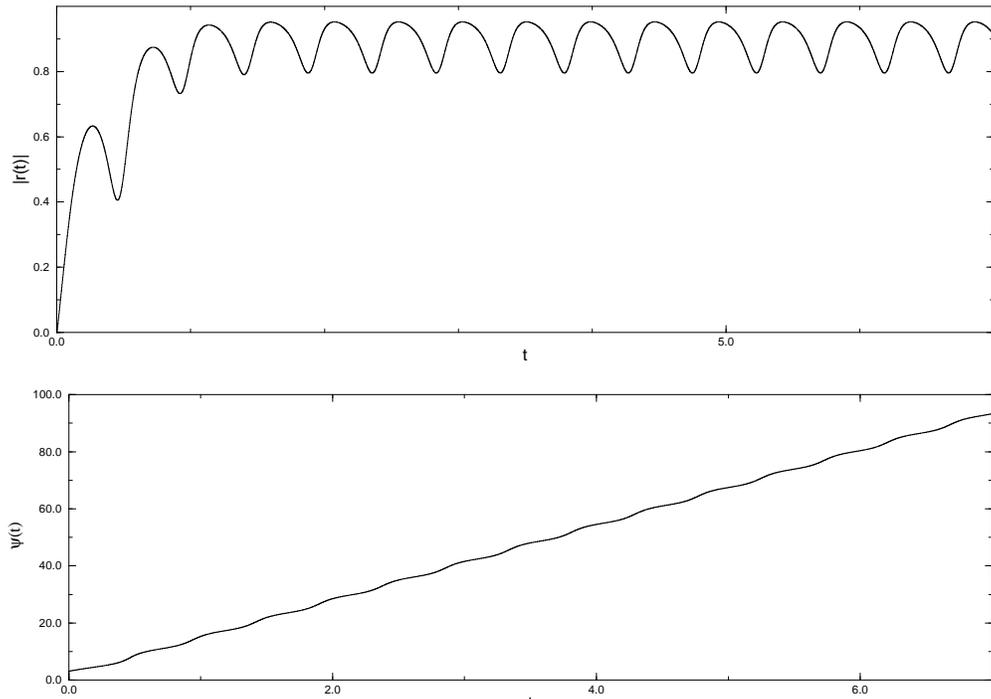,width=10.0cm,angle=-90}}}
\caption{Same as Fig.~7 for a larger $h_0$, $h_0= 7.5$. The oscillations 
of the order parameter amplitude become more pronounced. Only simulation
data are depicted.}
\label{fig8}
\end{figure}

\begin{figure}
\centerline{\hbox{\psfig{figure=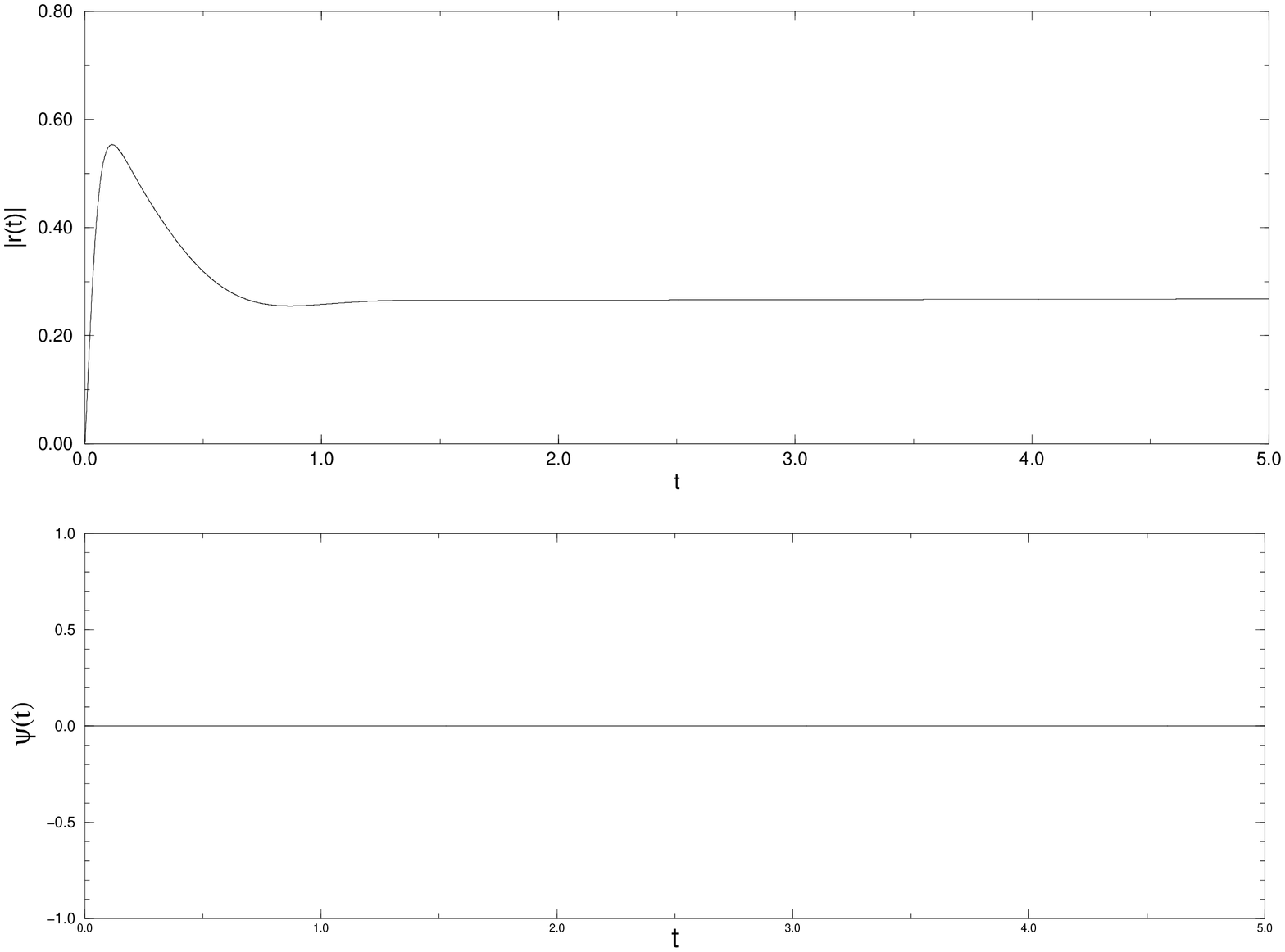,width=10.0cm,angle=-90}}} 
\caption{Same as Fig.~7 for the following parameter values: $\omega_0 = h_0 =
20$, and $K=6$. Notice that for such large values of the external field,
a stationary state is reached for long times.}
\label{fig9}
\end{figure}

Our asymptotic method yields analytical results when external fields 
of magnitude small compared to $\omega_0$ are included. For the sake 
of simplicity we shall present results corresponding to unimodal 
frequency distributions, $g(\omega) = \delta(\omega-\omega_0)$, and 
external field distributions, $f(h) = \delta(h-h_0)$. Then the probability
density has a unique component rotating at frequency $\omega_0$ which 
evolves toward the stationary distribution (\ref{mm15}) (in the comoving 
frame). This prediction is qualitatively supported by the numerical 
simulations as depicted in Fig.~\ref{fig7}. The numerical results show 
that the amplitude of the order parameter oscillates about the constant
value predicted by our asymptotics. The difference is of order $\varepsilon$
and it could be accounted by the first correction to the leading-order result.
Fig.~\ref{fig8} shows that the oscillation of the order parameter amplitude
is enhanced as $h_0$ increases. Finally all oscillations disappear if 
the external field becomes of the same order as the frequency, as 
depicted in Fig.~\ref{fig9}. Notice that our method supports [in the
limit $\omega_0\to\infty$, $h=O(1)$] a conjecture by Arenas and P\'erez 
Vicente~\cite{ARE2}: the amplitude of the order parameter in the 
oscillatorily synchronized state (in the presence of an external field) 
is given by the same expression as in the stationary state if the exact 
time-dependent phase of the order parameter is inserted (instead of the
stationary phase). Of course it seems that the conjecture holds for a wide 
variety of parameter values, some outside the range of validity of our 
asymptotic method~\cite{ARE2}. 

\section{Conclusions}
The high-frequency limit of the mean-field Kuramoto-Sakaguchi model 
of oscillator synchronization has been studied by new multiscale 
and numerical spectral methods. The main result of the multitime scale 
method is that the probability density splits into independent components 
corresponding to the different peaks in the oscillator frequency distribution.
Each density component evolves towards a stationary distribution in a 
comoving frame rotating with the frequency of the corresponding peak 
in the oscillator frequency distribution. The overall order parameter 
may be calculated by putting together the partial order parameters of
different components. This gives a simple picture of overall oscillatory 
synchronization by studying synchronization of each density component. 
Our method gives the same results as bifurcation theory for those 
parameter values where both approximations hold. Our asymptotic method also
works far from bifurcation points and it agrees well with results of 
numerical simulations. We have used a spectral method to numerically
integrate the nonlinear Fokker-Planck system for frequency values where
simple finite differences break down. This method has merit in itself and
should be studied in more detail separately.


\section{ACKNOWLEDGMENTS}

We are indebted to F.\ Ritort and R.\ Spigler for fruitful discussions.  
We acknowledge financial support from the the Spanish DGICYT
through grant PB95-0296.


\begin{references} 
\bibitem[*]{acebron:email} {E-address \tt acebron@dulcinea.uc3m.es}

\bibitem[\dagger]{bonilla:email} {E-address \tt bonilla@ing.uc3m.es}. Author to
whom all correspondence should be addressed.

\bibitem{TS1} K.Y. Tsang, S.H. Strogatz, and K. Wiesenfeld, {\it Reversibility
and noise sensitivity of Josephson arrays}, Phys. Rev. Lett. {\bf 66}, (1991),
1094-1097.

\bibitem{TS2} K.Y. Tsang, R.E. Mirollo, S.H. Strogatz, and K. Wiesenfeld, {\it
Dynamics of a globally coupled oscillator array}, Physica D {\bf 48}, (1991),
102-112.

\bibitem{CDW} S.H. Strogatz, C.M. Marcus, R.M. Westervelt and R.E.
Mirollo, {\it Simple model of collective transport with phase slippage}.
Phys. Rev. Lett. {\bf 61}, (1988) 2380-2383.

\bibitem{WIN} A.T. Winfree, {\it Biological rhythms and the behavior of
populations of coupled oscillators}, J. Theoret. Biol. {\bf 16}, (1967), 15-42.

\bibitem{STRO} S. H. Strogatz, {\it Norbert Wiener's brain waves}, edited
by S. Levin. Lect. N. Biomath. {\bf 100}, Springer, N. Y. 1994.

\bibitem{MISTRO} R. E. Mirollo and S. H. Strogatz, {\it Synchronization of the
pulse-coupled biological oscillators}, SIAM J. Appl. Math. {\bf 50}, (1990),
1645-1662.

\bibitem{GRAY} C.M. Gray and W. Singer, {\it Stimulus specific neuronal 
oscillations in the cat visual cortex: a cortical functional unit}, Soc. 
Neurosci. Abst. {\bf 13}, (1987) 13.

\bibitem{KURAM2} Y. Kuramoto, {\it Chemical Oscillations, Waves and
Turbulence}. Springer, N. Y. 1984.

\bibitem{ARO} D. G. Aronson, G. B. Ermentrout and N. J. Kopell, {\it Amplitude
response of coupled oscillators}, Physica D {\bf 41}, (1990) 403-449.

\bibitem{BON1} L. L. Bonilla, {\it Stable Probability Densities and Phase 
Transitions for Mean-Field Models in the Thermodynamic Limit}, 
J. Statist. Phys. {\bf 46}, (1987) 659-678.

\bibitem{STROMI} S.H. Strogatz and R.E. Mirollo, {\it Stability of incoherence
in a population of coupled oscillators}, J. Statist. Phys. {\bf 63}, (1991), 
613-635.

\bibitem{BNS} L. L. Bonilla, J. C. Neu, and R. Spigler, {\it Nonlinear
stability of incoherence and collective synchronization in a population of
coupled oscillators}, J. Statist. Phys. {\bf 67}, (1992), 313-330.

\bibitem{CRAW} J.D. Crawford, {\it Amplitude expansion for instabilities in
populations of globally-coupled oscillators}, J. Statist. Phys. {\bf 74},
(1994), 1047-1084.

\bibitem{OKUDA} K. Okuda and Y. Kuramoto, {\it Mutual entrainment between
populations of coupled oscillators}, Prog. Theor. Phys. {\bf 86},
(1991), 1159-1176.

\bibitem{BPVS} L. L. Bonilla, C. J. P\'erez Vicente, and R. Spigler, {\it 
Time-periodic phases in populations of nonlinearly coupled
oscillators with bimodal frequency distributions}, Physica D  (1996), 
submitted.

\bibitem{KURAM}  Y. Kuramoto, {\it Self-entrainment of a population of coupled
nonlinear oscillators}; in {\it International Symposium on Mathematical
Problems in Theoretical Physics}, H. Araki ed., Lecture Notes in Physics {\bf
39}, Springer, N. Y. 1975. pp.\ 420-422.

\bibitem{SAK} H. Sakaguchi, {\it Cooperative phenomena in coupled
oscillator systems under  external fields}. Prog. Theor. Phys. {\bf 79},
(1988) 39-46. 

\bibitem{STROMI2} S. H. Strogatz and R. E. Mirollo, {\it Phase-locking and 
critical phenomena in lattices of coupled nonlinear oscillators with random 
intrinsic frequencies}. Physica D {\bf 31}, (1988) 143-168.

\bibitem{LUMER} E. D. Lumer and B. A. Huberman, {\it Hierarchical dynamics in 
large assemblies of interacting oscillators}. Phys. Lett. A {\bf 160}, (1991)
 227-230.

\bibitem{BON2} L. L. Bonilla and J. M. Casado, {\it Dynamics of a soft-spin 
van Hemmen model. I. Phase and bifurcation diagrams for stationary 
distributions}. J. Statist. Phys. {\bf 56}, (1989) 113-125. 

\bibitem{PAB} C. J. P\'erez Vicente, A. Arenas and L.\ L.\ Bonilla, {\it
On the short time dynamics of networks of Hebbian coupled oscillators}. J.
Phys. A {\bf 29}, (1996) L9-L16.

\bibitem{BON3} L.L. Bonilla, C.J. P\'erez Vicente and J.M. Rub\'{\i}, {\it
Glassy synchronization in a population of coupled oscillators}. J. Stat.
Phys. {\bf 70}, (1993) 921-937.

\bibitem{SOMP} H. Sompolinsky, D. Golomb and D. Kleinfeld, {\it Cooperative 
dynamics in visual processing}. Phys. Rev. A
{\bf 43}, (1991) 6990-7011.

\bibitem{ARE1} A. Arenas and C. J. P\'erez Vicente, {\it Phase diagram of 
a planar XY model with random field}. Physica A
{\bf 201}, (1993) 614-625.

\bibitem{ARE2} A. Arenas and C. J. P\'erez Vicente, {\it Exact long-time 
behavior of a network of phase oscillators under random fields}. Phys. Rev.
Lett. {\bf 50}, (1994) 949-952.

\bibitem{PR}
 C. J. P\'erez and F. Ritort, {\it A new approach to the dynamical solution 
of the Kuramoto model}. Preprint, Dec. 13-1996.

\bibitem{KELLER} 
R. M. Lewis and J. B. Keller, {\it Solution of the functional equation 
for the statistical equilibrium of a crystal}. Phys. Rev. {\bf 121}, (1961) 
1022-1037.

\bibitem{HOPF} 
E. Hopf, {\it Statistical hydromechanics and functional calculus}. J. Rat.
Mech. Anal. {\bf 1}, (1952) 87-123.

\bibitem{KRAI} 
R. M. Lewis and R. H. Kraichnan,  {\it A space-time functional formalism for 
turbulence}. Comm. Pure Appl. Math. {\bf 15}, (1962) 397-411.




\end{references}
\end{document}